\documentstyle [12pt,epsf] {article}

\catcode`@=12
\newcommand{\be}{\begin{equation}}
\newcommand{\ee}{\end{equation}}
\newcommand{\ber}{\begin{eqnarray}}
\newcommand{\eer}{\end{eqnarray}}
\newcommand{\bers}{\begin{eqnarray*}}
\newcommand{\eers}{\end{eqnarray*}}
\newcommand{\bt}{\begin{itemize}}
\newcommand{\et}{\end{itemize}}

\begin{document}
\vspace{0.5in}
\oddsidemargin -.375in
\newcount\sectionnumber
\sectionnumber=0
\def\bra#1{\left\langle #1\right|}
\def\ket#1{\left| #1\right\rangle}
\def\be{\begin{equation}}
\def\ee{\end{equation}}
\thispagestyle{empty}
\begin{flushright} UdeM-GPP-TH-01-93 \\
UTPT-01-13\\
ANL-HEP-PR-01-109\\
TAUP 2680-01\\
November 2001\
\end{flushright}
\vspace {.5in}
\begin{center}
{\Large\bf { Simple 
 Relations for two body B
 Decays to Charmonium and tests for $\eta-\eta^{\prime}$ mixing.}  \\}
\vspace{.5in}
{{\bf Alakabha Datta{\footnote{email: datta@lps.umontreal.ca}} 
${}^{a}$}, {\bf Harry J. Lipkin {\footnote{email: 
harry.lipkin@weizmann.ac.il}} 
${}^{b}$} and
{\bf Patrick. J. O'Donnell {\footnote{email:
pat@medb.physics.utoronto.ca}}${}^{c}$} 
 \\}
\vspace{.1in}
${}^{a}$
{\it Laboratoire Ren\'e J.-A. L\'evesque, Universit\'e de
Montr\'eal,} \\
{\it C.P. 6128, succ.\ centre-ville, Montr\'eal, QC, Canada H3C 3J7} \\
${}^{b)}$ {\it
Department of Particle Physics,\\
Weizmann Institute,\\
Rehovot 76100, Israel \\and\\
School of Physics and Astronomy, \\
Raymond and Beverly Sackler Faculty of Exact Sciences\\
Tel-Aviv University,\\
Tel-Aviv, Israel \\
and\\
High Energy Physics Division\\
 Argonne National Laboratory\\
Argonne, IL 60439-4815, USA\\
${}^{c}$ {\it Department of Physics and Astronomy,\\
University of Toronto, Toronto, Canada.}\\
}
\end{center}

\begin{abstract}
The two body decays of $B_d$ and $B_s$ decays into $J/\psi M$, where 
$M$ is a light meson, is studied
under the very simple assumptions that  the spectator quark does
not play a role in the decay of the weak heavy quark or antiquark.
This hypothesis leads to interesting relations between decay amplitudes.
The  assumption of $SU(3)$ symmetry leads to additional relations 
between the decay amplitudes and in particular,
the eight 
CP eigenstates $J/\psi K_S$, $J/\psi
\eta$, $J/\psi \eta^{\prime}$ and $J/\psi \pi^o$ are all 
given in terms of three
parameters. 
If agreement with experiment validates these assumptions the
parameters over determined by the results will give information about the ratio
of penguin to tree contributions to the "golden channel" $B^o \to J/\psi K_S$
decay and will provide 
 tests for the standard $\eta-\eta^{\prime}$ mixing, 
which assumes that this mixing is determined by a single
mixing angle, as well as determine the value of the mixing angle.
We also present tests of the standard $\eta-\eta^{\prime}$ mixing 
involving semileptonic $D$ decays.
\end{abstract}

\newpage \pagestyle{plain}

\section{Introduction - The inactive spectator approach}

New data will soon accumulate on both $B^o$ and $B_s$ decays into  final CP
eigenstates containing charmonium. We present a simple method to facilitate
their  analysis and the extraction of parameters relevant to CP violation.
These states are of particular interest both because they provide ``golden
channels" like $J/\psi K_S$ important for CP  violation, and because the
presence of a bound $c \bar c$ pair greatly simplifies the analysis of large
groups of different decays related by symmetries. The $c \bar c$ pair is a
singlet under color, isospin and flavor SU(3) and is an eigenstate of $C$ and
$P$. The color coupling of the two-meson  final state is unique and the $c \bar
c$ pair is inert under the various symmetries which act only on the light quark
pair. 

The dominant tree and penguin diagrams  describing nonleptonic  $B_d $ and
$B_s$ decays to charmonium and a meson can all be described as a $\bar b$ decay
in which the spectator quark does not participate in a flavor-changing
interaction and later combines with a light antiquark to make the final 
light  meson. We now apply this 
``inactive spectator" approach to all such decays and first note a selection 
rule that forbids all  decays in which the spectator quark does not appear in
the final state:
\ber
 A[ {B}^0 \to J/\psi M(\bar q s)]  =0 \nonumber\\ A[ {B}_s \to J/\psi M(\bar
q d)]  =0  
\label{spectat} 
\eer	  
where $M(\bar q s)$ and $M(\bar q d)$
denote respectively any $\bar q q$ meson  meson with the constituents $\bar q
s$ and $\bar q d$. This selection rule can immediately be tested in many ways
to check the validity of our basic assumption when data are available; e.g. 
\ber
A_L( {B}^0 \to J/\psi \rho^o)  &=& A_L( {B}^0 \to J/\psi \omega) 
\label{spectatest1} 
\eer	      
\ber 
A_L( {B_s} \to J/\psi \rho^o)  &=&
A_L( {B_s} \to J/\psi \omega) = A_L( {B}^0 \to J/\psi \phi) = 0\
\label{nuspectest2} \eer 
where $L$ denotes any partial wave for the vector-vector final state in any
basis; e.g. in the s, p, d orbital angular momentum basis, the helicity basis 
or the transversity basis.

Eq.~(\ref{spectatest1}) is a particularly robust test for violations of our
approach by the presence of contributions in which the spectator quark is
annihilated. The $\rho^0$ and $\omega$ have opposite  relative signs in the
forbidden $u{\bar{u}}$ and allowed $d{\bar{d}}$ components of their
wavefunctions. The branching ratios of ${B_d}$ to $J/\psi \rho^o$ and  $J/\psi
\omega$ can differ by a factor of 2 if the forbidden $u{\bar{u}}$ amplitude is
$20\%$ of the allowed $d{\bar{d}}$ amplitude, which means only a $4\%$ ratio of
the direct forbidden to direct allowed contributions. Such forbidden  
contributions can arise from  diagrams expected to be small; e.g.  $W$
exchange  OZI violating diagrams of the type  $ \bar{b} +d=\bar{u} + u
+3G=\bar{u}+ u +J/\psi $.   There can also be contributions of the type
$\bar{b} +d=\bar{c} + c +3G=J/\psi +V $ which  violate Eq.~\ref{nuspectest2}.
Whether these contributions are indeed small can now be checked by
experimental tests of eqs.~(\ref{spectatest1}-\ref{nuspectest2}) and sensitive
upper limits on their magnitudes can be given if they are not observed..

If the absence of these transitions is confirmed, the remaining decays are all
describable by the two transitions: 
\ber 
B(\bar b q) \to J/\psi \bar d q  & 
\to & J/\psi M(\bar d q) \nonumber\\ B(\bar b q) \to J/\psi \bar s q  &  \to
& J/\psi M(\bar s q)  
\label{bquarkdecay} 
\eer 
where $B(\bar b q)$ denotes a $B$ meson with the  quark constituents $\bar b q$,
the spectator quark $q$ can be $s$, $u$ or $d$, and $M(\bar d q)$ and
$M(\bar sq)$  denote the final mesons.

The decay amplitudes are then described as the product of a $\bar b$ decay
amplitude and a hadronization function h describing the combination of a 
quark-antiquark pair to make the final meson.
\ber
A[ {B}^0 \to J/\psi M^0(\bar s d)] 
 &=& A( {\bar b} \to J/\psi \bar s) \cdot  h[\bar s d \to  M^0(\bar s d)] \nonumber\\
A[ {B}_s \to J/\psi M^0(\bar d s)] 
 &=& A( {\bar b} \to J/\psi \bar d) \cdot  h[\bar d s \to  M^0(\bar d s)] 
  \label{Bdecays}
\eer	     
\ber
A[ {B_d}^0 \to J/\psi M^0(\bar d d)] 
 &=& A( {\bar b} \to J/\psi \bar d) \cdot  h[\bar d d \to  M^0(\bar d d)] \nonumber\\
A[ {B}_s \to J/\psi M^0(\bar s s)] 
 &=& A( {\bar b} \to J/\psi \bar s) \cdot  h[\bar s s \to  M^0(\bar s s)]
  \label{Bdecayn}
\eer	     
\ber
A[ {B}^+ \to J/\psi M^+(\bar s u)]
 &=& A( {\bar b} \to J/\psi \bar s) \cdot  h[\bar s d \to  M^+(\bar s u)] \nonumber\\
A[ {B}^+ \to J/\psi M^+(\bar d u)]
 &=& A( {\bar b} \to J/\psi \bar d) \cdot  h[\bar d d \to  M^+(\bar d u)]
  \label{Bdecay+}
\eer
where the relations apply for any charmonium state as well as $J/\psi$.
The charged $B^+$ decays (\ref{Bdecay+}) are uniquely related by isospin
symmetry to the corresponding $B^0$ decays and are not considered further.  

The pairs of decays (\ref{Bdecays}) into charge-conjugate strange final states
are related by charge conjugation invariance which is valid for all strong 
interactions. They differ only by the weak interaction vertex which violates
$C$ and by possible kinematic and form  factor differences induced by the  $B_d
- B_s$ mass difference. For instance, the hadronization functions for  these
decays, which have the same functional dependence because of  charge
conjugation,  depend on the magnitude of the relative momentum of the s and 
the d quark which are the same in both decays up to corrections from the
$B_s-B_d$ mass difference.  Hence the values of the  hadronization functions
for the two decays are the same up to corrections  from the $B_s-B_d$ mass
difference.     Note also that there is no $SU(3)$ breaking in final state
interactions for the above final states because of charge conjugation
symmetry\cite{Lipkin1997}. This is true for elastic as well as inelastic
rescattering where an  example of the latter are the processes $B_d \to
D^-D_s^{+} \to J/\psi K^{*0}$ and $B_s \to D^+D_s^{-} \to J/\psi \bar K^{*0}$.
Note that $SU(3)$ breaking  may arise because of differences in the production
of the $D^-D_s^{+}$ and $D^+D_s^{-}$ intermediate states but the transition
from the intermediate state to the final state are equal in the two decays
because of charge conjugation symmetry.

The decays into strange (\ref{Bdecays}) and nonstrange (\ref{Bdecayn}) final
states are related by SU(3) symmetry which in this model only affects the
hadronization functions which differ by interchanging $s$ and $d$ flavors and 
possible kinematic and form  factor differences induced by the  mass
differences.

\section{B decays into charmonium and a vector meson}

We first consider the decays $A(B \to J/\psi V)$ where $V$ is a vector meson 
and immediately  obtain the selection rules
(\ref{spectatest1}-\ref{nuspectest2}). Next we note that charge conjugation
invariance, as already discussed  above, requires that for a given  partial
wave \cite{Lipkin1997}
\be
 A( {B_s} \to J/\psi  \bar K^{*0})_L = F_{CKM}^{L}\cdot
A( {B_d} \to J/\psi  K^{*0})_L 
\label{cc} 
\ee
where $ F_{CKM}^L$ is a factor depending on the ratios of the $CKM$ matrix
elements and the ratio of various weak interaction diagrams; e.g. penguin and
 tree, contributing to the $B_d$ and $B_s$ decays. 
\ber
F_{CKM}^{L}& = &\frac
 {A_L ( {\bar b \to J/\psi  \bar d)}}
{A_L( {\bar b \to J/\psi \bar s)}}\
\label{FCKM} 
\eer
For the dominant tree diagram and penguin diagram contributions with a charmed
quark loop the weak transition is 
$\bar b \to \bar c + W^+ \to \bar c + c + \bar q$
where $q$ is $d$ or $s$. For this case  $ F_{CKM}^L= V_{cd}/V_{cs}$.

Finally, the additional assumption of SU(3) symmetry leads to the full
set of predictions
\ber
A_L( {B_d}^0 \to J/\psi \rho^o) 
 &=& A_L( {B_d}^0 \to J/\psi \omega) = (1/{\sqrt 2})\cdot  
 A_L( {B_s} \to J/\psi  \bar K^{*0}) 
 \nonumber\\
A_L( {B}_s \to J/\psi \phi) 
 &=& A_L( {B}^0 \to J/\psi  K^{*0})
  \nonumber\\
A_L( {B_s} \to J/\psi \rho^o) 
 &=& A_L( {B_s}^0 \to J/\psi \omega) = A_L( {B}^0 \to J/\psi \phi) = 0
\
\label{vectorsu3}
\eer

\section{$\eta-\eta^{\prime}$ mixing and B decays into charmonium and a 
pseudoscalar meson}

We now note that Eq.~\ref{Bdecayn} leads to relations between amplitudes
involving the $\eta$ or the $\eta^{\prime}$ in the final state. These decays
are interesting as recent experimental data for $B$  decays into such final 
states have so far remained unexplained by the standard treatments of these
decays \cite{Lipkin2000}.

The most general description of $\eta-\eta^{\prime}$ system involves 
four different radial
wave functions and cannot be described by diagonalizing a simple 
$2 \times 2$ matrix with a single mixing angle. One can therefore write the
normalized $\eta-\eta^{\prime}$ wavefunctions as
\ber
\ket{\eta} & = & \cos{\phi}\ket{N} -\sin{\phi}\ket{S}\nonumber\\
\ket{\eta^{\prime}} & = & \sin{\phi^{\prime}}\ket{N^{\prime}} +
\cos{\phi^{\prime}}\ket{S^{\prime}}\
\label{general}
\eer
where $\ket{N}$, $\ket{N^{\prime}}$, $\ket{S}$ and $\ket{S^{\prime}}$ are respectively
arbitrary isoscalar nonstrange and strange quark-antiquark wavefunctions.
In the traditional picture, where the $\eta-\eta^{\prime}$ mixing is described by a single mixing angle,
\ber
\ket{N} & = & \ket{N^{\prime}}\nonumber\\
\ket{S} & = & \ket{S^{\prime}}\nonumber\\
\phi & = & \phi^{\prime}\
\label{std}
\eer 
A particular example of the general $\eta-\eta^{\prime}$ mixing can be found 
in Ref\cite{DLO}
where  we  considered
the possibility that the $\eta$ and $\eta^{\prime}$ 
wavefunctions are mixtures of ground state and radially excited
$q \bar q$ systems.  

Note that in the $B_d$ decays the $\eta$ and the $\eta^{\prime}$ are produced via their
nonstrange components,
$ B_d \to J/\psi N(N^{\prime}) \to J/\psi \eta(\eta^{\prime})$  
while in $B_s$ decays
the $\eta$ and the $\eta^{\prime}$ are produced via their strange components,
$ B_s \to J/\psi S(S^{\prime}) \to J/\psi \eta(\eta^{\prime})$.
With the mixing in Eq.~\ref{general} and Eq.~\ref{Bdecayn} we get
the following predictions.
\ber
A(B_d \to J/\psi \eta) & = & A(B_d \to J/\psi N)\cos{\phi}\nonumber\\
A(B_d \to J/\psi \eta^{\prime}) & = & A(B_d \to J/\psi N^{\prime})
\sin{\phi^{\prime}}\nonumber\\
A(B_s \to J/\psi \eta) & = & -A(B_s \to J/\psi S)\sin{\phi}\nonumber\\
A(B_s \to J/\psi \eta^{\prime}) & = & A(B_s \to J/\psi S^{\prime})
\cos{\phi^{\prime}}\
\label{amp}
\eer
With the the standard mixing in Eq.~\ref{std} we find
\ber
A(B_d \to J/\psi \eta)=\cot{\phi} \cdot A(B_d \to J/\psi \eta^{\prime})\nonumber\\
A(B_s \to J/\psi \eta)=-\tan{\phi} \cdot A(B_s \to J/\psi \eta^{\prime})\
\label{std1}
\eer
These relations were obtained in Ref\cite{Lipkin1997} including only 
tree diagrams. However, these relations continue to be true even when
a penguin contribution is included 
 as long as the model of nonleptonic decays in 
Eq.~\ref{Bdecayn} is valid.

We thus see that the mixing angle with standard mixing can be obtained from
experiment in two different ways. We can then test standard mixing  by seeing
that both ways give the same result. Allowing for the $\eta-\eta^{\prime}$ mass
difference and including phase space factors we can construct ratios of
experimentally measured quantities,

\ber
r_d & \equiv &\frac
{p_{\eta^{\prime}}^3\Gamma( {\bar B}^0 \to J/\psi \eta) }
{p_{\eta}^3\Gamma( {\bar B}^0 \to J/\psi \eta^{\prime})} = \cot^2\phi\
\eer

\ber
r_s & \equiv &\frac
{p_{\eta^{\prime}}^3\Gamma( {\bar B_s}^0 \to J/\psi \eta) }
{p_{\eta}^3\Gamma( {\bar B_s}^0 \to J/\psi \eta^{\prime})} = \tan^2\phi\
\eer
We then have the prediction
\ber
r & = & \sqrt{r_dr_s}=1\
\label{rnl}
\eer 
Any large deviation of $r$ from 1 would indicate evidence of non standard
$\eta-\eta^{\prime}$ mixing.
To see what we might expect for the values of $r$  with 
non standard 
mixing we consider the example of non standard mixing considered in 
Ref\cite{DLO} where we find that
 the ratio $r$ can be in the range $r=0.82 - 0.2$. Hence, 
in general,  
deviation of $r$ from unity by a factor of 2 or more 
would be an unambiguous
signal for nonstandard $\eta-\eta^{\prime}$ mixing.
In light of our earlier discussion we note that the relations
in Eq.~\ref{amp}
  will be violated  in 
  $W$ exchange OZI violating diagrams in
$B_s$ decay of the type $ \bar{b} +s=\bar{u} + u +3G=\bar{u}+ u +J/\psi $.
One can also have OZI violating diagrams
due to the anomaly, gluon couplings to the flavor singlet 
component of the $\eta^{\prime}$ or the intrinsic charm content of 
the $\eta(\eta^{\prime})$ 
\cite{Atwood:1997bn, Halperin:1998ma} leading to the processes
$ \bar{b} +s(d)=\bar{c} + c +2G=J/\psi +\eta^{\prime} $ 
and
$ \bar{b} +s(d)=\bar{c} + c +3G=J/\psi +\eta(\eta^{\prime}) $
that will violate the relations in Eq.~\ref{amp}.
Hence the violation of the prediction in Eq.~\ref{amp} and Eq.~\ref{rnl} 
would indicate evidence of non standard $\eta-\eta^{\prime}$ mixing
and/or the presence of OZI violating contributions.
 
Unfortunately, the large $\eta-\eta^{\prime}$ mass differences may introduce 
other corrections beyond simple phase space.
However, we can extend our tests of standard mixing by noting that the
assumptions that the states $\ket{N}$ and $\ket{S}$ have the same radial
wave functions implies an SU(3) symmetry that also includes the kaon wave
functions. We can therefore include the transitions to final states with kaons
 produced by the same $\bar b  \to
 J/\psi + \bar d$ or $\bar b  \to J/\psi  + \bar s$ decays and differing
 only by the flavor of the spectator quark. We then have the predictions,
\be
{\sqrt 2} \cdot A ({B_d} \to J/\psi N)  =  A( {B_s} \to J/\psi  \bar K^0)
={\sqrt 2}A( {B_d} \to J/\psi   \pi^0)\
\ee
\be
A( {B_s} \to J/\psi S) = 
A( {B_d} \to J/\psi  K^0) 
\ee

These can be combined with Eq.~\ref{amp} to give sum rules independent of the
mixing angle for standard mixing,

\ber
|A(B_d \to J/\psi \eta)|^2 +
|A(B_d \to J/\psi \eta^{\prime})|^2 & = & 
(1/2)\cdot |A( {B_s} \to J/\psi  \bar K^0)|^2\nonumber\\
|A(B_s \to J/\psi \eta)|^2 +
|A(B_s \to J/\psi \eta^{\prime})|^2 & = & 
|A( {B_d} \to J/\psi  K^0)|^2
\
\label{sumrule}
\eer

As in Eq.~\ref{cc} charge conjugation requires
\be
 A( {B_s} \to J/\psi  \bar K^0) = F_{CKM}\cdot
A( {B_d} \to J/\psi  K^0) 
\ee
where $ F_{CKM}$ is defined as in Eq.~\ref{FCKM} but for the non spin 
flip transition needed to form a spin zero meson in the final state.
\ber
F_{CKM}& = &\frac
 {A ( {\bar b \to J/\psi  \bar d)}}
{A( {\bar b \to J/\psi \bar s)}} 
\eer

Thus to  obtain a different point of  view we can define the ratios of $B_d$ and
$B_s$ decays to $J/\psi \eta$ and $J/\psi \eta^{\prime}$.
\ber
r_\eta &  = &\frac
{p_{B_s\eta}^3\Gamma( {B_d} \to J/\psi \eta) }
{p_{B_d\eta}^3\Gamma({B_s} \to J/\psi \eta)} = (F_{CKM})^2\cdot\cot^2\phi\
\eer
\ber
r_\eta^{\prime}& = &\frac
{p_{B_s\eta^{\prime}}^3\Gamma( {\bar B}^0 \to J/\psi \eta^{\prime}) }
{p_{B_d\eta^{\prime}}^3\Gamma( {\bar B_s} \to J/\psi \eta^{\prime})} = 
( F_{CKM})^2\cdot \tan^2\phi\
\eer
We then have the prediction
\ber
r_B & = & \sqrt{r_\eta r_\eta^{\prime}}= (F_{CKM})^2\
\label{rn2}
\eer 
We can refine these ratios and overcome kinematic factors by normalizing them 
to the kaon modes,
\ber
R_\eta & = &\frac
{p_{B_s\eta}^3\Gamma( {B}^0 \to J/\psi \eta) }
{p_{B_d\eta}^3\Gamma( {B_s} \to J/\psi \eta)} \cdot
\frac
{p_{BdK}^3\Gamma( {B_s} \to J/\psi  K^0)}
{p_{BsK}^3\Gamma( {B_d} \to J/\psi \bar K^0) }
= (F_{CKM})^4\cdot \cot^2\phi\
\eer
\ber
R_\eta^{\prime}& = &\frac
{p_{B_s\eta^{\prime}}^3\Gamma( {B}^0 \to J/\psi \eta^{\prime}) }
{p_{B_d\eta^{\prime}}^3\Gamma(B_s) \to J/\psi \eta^{\prime})} 
\cdot
\frac
{p_{BdK}^3\Gamma( {B_s}^0 \to J/\psi  K^0)}
{p_{BsK}^3\Gamma( {B}^0 \to J/\psi \bar K^0) }
= 
(F_{CKM})^4\cdot \tan^2\phi\
\eer
We then have the prediction
\ber
R_B & = & \sqrt{R_\eta R_\eta^{\prime}}=(F_{CKM})^4\
\label{rn3}
\eer 

We now note that under the assumptions of standard mixing, SU(3) symmetry and
the non-participation of the spectator quark in the weak transition we have
described the branching ratios for eight transitions in terms of three
parameters, $F_{CKM}$, $\phi$ and an overall normalization. If these relations
hold experimentally, the standard mixing and the value of the mixing angle will
be confirmed and established, the validity of SU(3) symmetry for these
transitions  will be confirmed, and the value of $F_{CKM}$ will determine the
ratio of the  penguin to tree contributions to the decay $B_d \to J/\psi  K_S$ 
which is the "golden channel" for CP violation experiments. If experimental
violations of this description are observed, they will indicate the breakdown
of particular assumptions and perhaps give clues to new physics. 

\section{$\eta-\eta^{\prime}$ mixing in charmed meson decays}

One could, in principle, construct similar ratios with $D(D_s) \to
\eta(\eta^{\prime}) P$ where  $P= \pi, \rho, K $. However, non nonleptonic 
$D$ decays are are not very well understood 
 \cite{lipkinclose} and therefore these decays are not 
very useful to test non standard $\eta-\eta^{\prime}$ mixing.
On the other hand semileptonic $D(D_s)$ decays of the type
$D(D_s) \to \eta(\eta^{\prime}) l \nu$ 
can provide clean tests for the $\eta-\eta^{\prime}$ mixing.

Here the lepton pair is a singlet like charmonium under all strong interaction
symmetries. But the lepton pair mass and final momentum have continuous
spectra, and effects of the $\eta-\eta^{\prime}$ mass difference  can be large.
We therefore must consider the decay dynamics in  more detail.
 
The  Lagrangian for the semileptonic $D(D_s)$ decays  
involving the transitions $c \rightarrow q l \nu$,
  where $q=s,d$ and $l= \mu, e$, has the standard current-current form 
after the $W$ boson is
integrated out in the effective theory.  
\begin{eqnarray} H_{W} & = &
\frac{G_F}{2 \sqrt{2}}V_{cq} {\bar q}\gamma_{\mu}(1-\gamma_{5}) c
{\bar{\nu}}\gamma^{\mu}(1-\gamma_{5})l \ \end{eqnarray} 
The differential decay distribution, neglecting the lepton masses, is 
then given by,
\ber
\frac{d\Gamma}{dq^2}(D_q \to P l \nu) & = &
\frac{G_F^2|V_{qc}|^2}{24 \pi^3}p^3_P|F_1^q(q^2)|^2\nonumber\\
p & = & \sqrt{E^2-m_P^2} \nonumber\\
E & = & \frac{m_D^2 +m_P^2-q^2}{2m_D}\
\eer
where $P=\eta(\eta^{\prime})$  and $p$ and $E$ are the magnitude of the
momentum and energy of the pseudoscalar meson $P$. 
 The form factor $F_1$ is defined as \cite{BSW}
\ber
\bra{P(p_f)}\bar{q} \gamma_\mu (1-\gamma_5) b \ket{ D_q(p_i) }
&=& \left[ (p_i+ p_{f})_\mu - \frac{m_{D_q}^2-m_{P}^2}{q^2} 
q_\mu \right]
F_1^q(q^2)\nonumber\\
& + & \frac{m_{D_q}^2-m_{P}^2}{q^2} q_\mu F_0^q(q^2)  \
\label{ffactor}
\eer
Let us now define the two ratios
\ber
r_{d} & = &\frac{\Gamma(D \to \eta l \nu)}{\Gamma(D \to \eta^{\prime} l \nu)} \nonumber\\
r_{s} & = &\frac{\Gamma(D_s \to \eta l \nu)}{\Gamma(D_s \to \eta^{\prime} 
l \nu)} \
\label{def_r}
\eer
It then follows from the mixing in Eq.~\ref{std} that 
\ber
r_d 
& = &
\frac{\int_{0}^{(m_D-m_{\eta})^2}p_{\eta}^3
|F_1^d(q^2)|^2dq^2}
     {\int_{0}^{(m_D-m_{\eta^{\prime}})^2}p_{\eta^{\prime}}^3
|F_1^d(q^2)|^2dq^2}
\cot^2{\phi}
\nonumber\\
r_s
& = &
\frac{\int_{0}^{(m_{D_s}-m_{\eta})^2}p_{\eta}^3
|F_1^s(q^2)|^2dq^2}
     {\int_{0}^{(m_{D_s}-m_{\eta^{\prime}})^2}p_{\eta^{\prime}}^3
|F_1^s(q^2)|^2dq^2}\tan^2{\phi}\
\label{rdrs}
\eer

To calculate $r_{d,s}$ we have to model the $q^2$ dependence of the 
form factors $F_1^{d,s}(q^2)$ and no simple observable - independent
of the form factors- can be constructed
that can test for nonstandard mixing even in the $SU(3)$ limit.

It is more useful to define the two ratios
\ber
r_{\eta} & = &\frac{\Gamma(D \to \eta l \nu)}{\Gamma(D_s \to \eta l \nu)} 
\nonumber\\
r_{\eta^{\prime}} & = &\frac{\Gamma(D \to \eta^{\prime} l \nu)}
{\Gamma(D_s \to \eta^{\prime} 
l \nu)} \
\label{ndef_r}
\eer
It then follows from the standard mixing in Eq.~\ref{std} that 
\ber
r_{\eta} 
& = &
\frac{\int_{0}^{(m_D-m_{\eta})^2}p_{\eta}^3
|F_1^d(q^2)|^2dq^2}
     {\int_{0}^{(m_{D_s}-m_{\eta})^2}p_{\eta}^3
|F_1^s(q^2)|^2dq^2}
\cot^2{\phi}
\nonumber\\
r_{\eta^{\prime}}
& = &
\frac{\int_{0}^{(m_{D}-m_{\eta^{\prime}})^2}p_{\eta^{\prime}}^3
|F_1^d(q^2)|^2dq^2}
     {\int_{0}^{(m_{D_s}-m_{\eta^{\prime}})^2}p_{\eta^{\prime}}^3
|F_1^s(q^2)|^2dq^2}\tan^2{\phi}\
\label{rdrs1}
\eer
In the U spin limit we then have the prediction
\ber
r_D & = & \sqrt{r_{\eta}r_{\eta^{\prime}}} =1\
\label{rsl}
\eer
Any large deviation of $r_D$ from 1 by a factor of 2 
would indicate evidence of non standard 
$\eta-\eta^{\prime}$ mixing as they are unlikely to 
originate from U spin breaking. 

With more experimental data one could  devise  tests of the
$\eta-\eta^{\prime}$ mixing by looking at the various decay distributions.
Let us  define the  two ratios
\ber
R_d(q^2) & = &\frac{p_{\eta^{\prime}}^3}{p_{\eta^3}}
 \frac{\frac{d\Gamma}{dq^2}(D_d \to \eta l \nu)}
                    {\frac{d\Gamma}{dq^2}(D_d \to \eta^{\prime} l \nu)}
\nonumber\\
R_s(q^2) & = &\frac{p_{\eta^{\prime}}^3}{p_{\eta^3}}
 \frac{\frac{d\Gamma}{dq^2}(D_s \to \eta l \nu)}
                    {\frac{d\Gamma}{dq^2}(D_s \to \eta^{\prime} l \nu)}\
\eer
It then follows for the standard mixing in Eq.~\ref{std} that
\ber
R_d(q^2) & = & \cot^2{\phi} \nonumber\\
R_s(q^2) & = & \tan^2{\phi} \
\label{pred1}
\eer
We see from the above equations that measurement of $R_{d,s}$ allows us 
to calculate the mixing angle $\phi$. What is also interesting is that the
the ratios $R_{d,s}(q^2)$ are independent of $q^2$ and we have the
prediction
\ber
R & = & R_d(q^2)R_s(q^2) =1\
\label{R}
\eer
for any value of $q^2$.
A deviation of $R$ from unity would indicate evidence of non standard mixing.
In particular, with the mixing in Ref\cite{DLO} one would predict a 
value of $R$ different from unity and a 
$q^2$ dependence for $R$, $R_d$ and $R_s$.

\section{Conclusions} 
 
In summary we studied the
 two body decays of $B_d$ and $B_s$ decays into $J/\psi M$, where 
$M$ is a light meson, 
under the very simple assumptions that  the spectator quark does
not play a role in the decay of the weak heavy quark or antiquark.
Using this model we derived
 interesting relations between decay amplitudes
and  tests for the standard $\eta-\eta^{\prime}$ mixing.
With the  further 
  assumption of $SU(3)$ symmetry we derived additional relations between
decay amplitudes 
 and in particular
the eight 
CP eigenstates $J/\psi K_S$, $J/\psi
\eta$, $J/\psi \eta'$ and $J/\psi \pi^0$ were given  in terms of three
parameters which, when determined from experiments,
 will give information about the ratio
of penguin to tree contributions to the "golden channel" $B^0 \to J/\psi K_S$
decay and will give the value of the $\eta - \eta^{\prime}$ mixing angle. 
We also presented tests of the standard $\eta-\eta^{\prime}$ mixing 
involving semileptonic $D$ decays.
  
\centerline{ {\bf  Acknowledgment}}
This work was  supported  by the
US-Israel Bi-National Science Foundation
and by the U.S. Department
of Energy, Division of High Energy Physics, Contract W-31-109-ENG-38
(Harry Lipkin)
 and by the Natural  Sciences and Engineering Research  Council
of Canada (A. Datta and P. J. O'Donnell).

\end{document}